\documentclass[12pt]{article}
\input epsf.sty
\usepackage{mathrsfs}
\usepackage{amsmath}
\usepackage{mathrsfs}
\usepackage{amssymb}
\usepackage{color}
\usepackage{psfrag}
\usepackage{graphicx}
\usepackage{subfigure}
\usepackage{rotating}

\newcommand{\bea}{\begin{eqnarray}}
\newcommand{\eea}{\end{eqnarray}}
\newcommand{\be}{\begin{equation}}
\newcommand{\ee}{\end{equation}}
\newcommand{\vs}[1]{\vspace{#1 mm}}

\newcommand{\dsl}{\pa \kern-0.5em /}

\newcommand{\half}{\frac{1}{2}}
\newcommand{\pa}{\partial}

\newcommand{\nn}{\nonumber\\}

\begin{document}
\topmargin 0pt
\oddsidemargin 0mm

\begin{flushright}


\end{flushright}

\vspace{2mm}

\begin{center}
{\Large \bf Zero sound in strange metals with hyperscaling violation from holography}  

\vs{10}

{Parijat Dey\footnote{E-mail: parijat.dey@saha.ac.in} and 
Shibaji Roy\footnote{E-mail: shibaji.roy@saha.ac.in}}

 \vspace{4mm}

{\em

 Saha Institute of Nuclear Physics,
 1/AF Bidhannagar, Calcutta-700 064, India\\}

\end{center}

\vs{10}

\begin{abstract}
Hyperscaling violating `strange metal' phase of heavy fermion compounds can be described
holographically by probe D-branes in the background of a Lifshitz space-time (dynamical
exponent $z$ and spatial dimensions $d$) with hyperscaling violation (corresponding exponent 
$\theta$). Without the hyperscaling violation, strange metals are known to exhibit zero sound
mode for $z<2$ analogous to the Fermi liquids. In this paper, we study its fate in the presence
of hyperscaling violation and find that in this case the zero sound mode exists for 
$z < 2(1+|\theta|/d)$, where the positivity of the specific heat and the null energy condition
of the background dictate that $\theta<0$ and $z\geq 1$. However, for $z \geq 2(1+|\theta|/d)$,
there is no well-defined quasiparticle for the zero sound. The systems behave like Fermi liquid
for $2|\theta|=dz$ and like Bose liquid for $2|\theta| = qdz$ (where $q$ is the number
of spatial dimensions along which D-branes are extended in the background space), but in 
general they behave
as a new kind of quantum liquid. We also compute the AC conductivity of the systems and
briefly comment on the results.        
      
\end{abstract}

\newpage
\noindent {\it 1.  Introduction} : 
AdS/CFT \cite{Maldacena:1997re} or more generally gauge/gravity duality \cite{Aharony:1999ti}
makes the strongly coupled field theories 
tractable by mapping it to a weakly coupled string or gravity theory living in one dimension 
higher. Among many examples, one class of systems in condensed matter is the strongly coupled 
quantum liquids \cite{Lifshitz, Abrikosov, Pines, Negele} where gauge/gravity duality may play 
important role to understand their 
unusual behavior at the theoretical level. Indeed, some of the thermodynamic and transport
properties of strange metals, like linear temperature behavior of DC resistivity and anomalous
power law tails of AC conductivity which can not be understood from perturbative Fermi liquid 
theory have been understood using gauge/gravity duality \cite{Hartnoll:2009ns}. The holographic 
model used for this
purpose is the probe D-brane in the background of Lifshitz space-time. While Lifshitz space-time
(with spatial dimensions $d$ and dynamical exponent $z$) is the holographic dual \cite{Koroteev:2007yp,
Kachru:2008yh} 
to a scale invariant, 
strongly interacting theory at quantum criticality, D-branes provide the finite density
charge carriers in the system \cite{Kobayashi:2006sb, Karch:2007pd, Karch:2007br} and they together 
show the characteristic strange metal behaviors \cite{Hartnoll:2009ns}.
Even though strange metals show non-Fermi liquid behavior (without any obvious Fermi surface),
it was shown holographically in \cite{HoyosBadajoz:2010kd} that there exists a zero sound\footnote{Zero
sound was first shown to exist from holography in \cite{Karch:2009zz} for D3-D7 system representing 
a new type of quantum liquid. This mode has also been shown to exist in some other holographic models
in \cite{Kulaxizi:2008kv}.} mode analogous to the
Fermi liquids \cite{Landau} when the dynamical exponent $z<2$. However, the mode gets washed 
away as we take
$z \geq 2$ and this is also corroborated by their results on AC conductivity in this model.

In this paper, we would like to understand the fate of the above mentioned zero sound mode in
the hyperscaling violating\footnote{Hyperscaling is a property of a system by which the
dimensionful quantities (for example, entropy) scale by their natural length dimension. When it 
is violated the scaling properties also change. Hyperscaling gets violated by the random field 
fluctuations (as shown in random field Ising model \cite{Fisher:1986zz}) over thermal fluctuations and therefore
the scaling properties of the entropy (say) also gets modified near the critical point.} strange
metals. In particular, we would like to see under what condition and in what parameter range the
zero sound mode (if it exists at all) survives when the hyperscaling violation is present.        
First of all, we note that the background geometry consisting of Lifshitz metric with hyperscaling
violation (with parameter $\theta$) is stable for $d-\theta>0$ \cite{Dong:2012se}. When the probe D-brane
is introduced, from the probe brane thermodynamics \cite{Karch:2009eb, Benincasa:2009be, Lee:2010uy}
we find that the specific heat of this system is positive
only when $\theta<0,\,z>0$ or $\theta >0,\,z <0$. However, the null energy condition \cite{Hoyos:2010at, Dong:2012se} of 
the background
dictates that $\theta<0$ and $z \geq 1$. So, only in this case we have a physically sensible gravity 
background which
is also stable. For such a strange metal the specific heat scales with temperature as $T^{2|\theta|/(dz)}$.
We thus find that for $2|\theta| = dz$, the system will behave like Fermi liquid whereas for $2|\theta|
=dzq$ (here $q$ being the number of spatial dimensions along which D-branes are extended in $d$-dimensional
background), the system will behave like Bose liquid \cite{Lee:2010uy} and for generic values it will 
behave as a new kind of
quantum liquid. We find that even though the specific heat in general has non-Fermi liquid behavior,
the system has a zero sound mode for $1\leq z<2(1+|\theta|/d)$ at zero temperature. In this case, the real 
part of the dispersion
curve is linear in momentum ($\omega = v_0 k$), whereas, the imaginary part has the dependence
$k^{2(1+|\theta|/d)/z}$. So, the system behaves like Fermi liquid, i.e., the imaginary parts of the spectral
curve goes as $k^2$ for $1+|\theta|/d =z$. Combining this with the behavior of the specific heat we find
that for $z=2$, both the specific heat and the zero sound have Fermi liquid behavior as predicted by
Landau \cite{Landau}. However, for $dz=2|\theta|$ (or $dzq=2|\theta|$) and $z \neq 1+|\theta|/d$, the specific heat 
behaves like Fermi (or Bose) liquid, but the zero sound does not behave like Fermi liquid.
On the other hand, for $dz \neq 2|\theta|$ and $z = 1 + |\theta|/d$, the specific heat does not behave
like Fermi liquid but the zero sound does.   
For generic values of $z$ we still have zero sound mode even though the specific heat shows a different
kind of quantum liquid behavior. However, for $z \geq 2(1+|\theta|/d)$, we do not find any quasiparticle
for the zero sound mode. 
      
We also compute the AC conductivity of the system. In this case we find that when $z < 2(1+|\theta|/d)$,
the AC conductivity has a frequency dependence $i/\omega$, which is the high frequency limit of the standard
Drude conductivity. For $z > 2(1+|\theta|/d)$, the AC conductivity has a frequency dependence which has an anomalous 
power law behavior given by $\omega^{-2(1+|\theta|/d)/z}$. This generalizes the results of \cite{HoyosBadajoz:2010kd}, 
when strange metals have hyperscaling violation.

\vspace{.5cm}

\noindent{\it 2. The holographic model} : The background we consider is the Lifshitz space-time with hyperscaling 
violation and the metric has the form \cite{Dong:2012se},
\be\label{hvlif}
ds^2_{d+2} = r^{\frac{2\theta}{d}}\left[-\frac{dt^2}{r^{2z}} + \frac{\sum_{i=1}^d (dx^i)^2}{r^2} + \frac{dr^2}{r^2}\right]
\ee
Note from above that the field theory, whose spatial dimension is $d$, lives on the boundary $r \to 0$. The 
boundary theory has a Lifshitz symmetry with hyperscaling violation and shows up in the metric \eqref{hvlif} as 
described in the following. The metric has apart from space-time translation symmetry, spatial rotation symmetry, a scaling 
symmetry $t \to \lambda^z t$, $x^i \to \lambda x^i$ and $r \to \lambda r$ upto a conformal factor, where $z$ is
the dynamical exponent. Due to the conformal
factor the full metric is not invariant under the scaling but changes as $ds_{d+2} \to \lambda^{\theta/d} ds_{d+2}$,
where $\theta$ is the hyperscaling violation exponent. We will assume that the above metric has been obtained 
from large $N$ limit of some brane configurations.  

Now to model strange metals with hyperscaling violation we introduce $N_f$ (we will assume that $N_f \ll N$
\cite{Karch:2002sh} such that D-branes do not affect the background space-time -- the probe approximation) 
number of coincident D-branes in
the background which in turn will introduce $N_f$ hypermultiplet fields propagating in $q+1$ dimensions
where $q$ is the spatial dimension\footnote{If $q<d$, then the field theory lives on $q$ spatial dimensions
otherwise $q$ must be equated to $d$ everywhere in what follows.} of the boundary theory. We set the scalar 
fields to zero and therefore the 
dynamics of the probe D-branes will be described by the $U(N_f)$-valued gauge field. We further restrict       
ourselves to $U(1) \subset U(N_f)$ gauge group and consider that only non-vanishing component of the gauge 
field is $A_t(r)$. 

The dynamics of the D-brane is given by the Abelian DBI action of the form,
\be\label{dbi}
\hat{S} = - N_f T_D V \int dr dt d^qx \sqrt{-{\rm det}\left[g_{ab} + (2\pi\alpha')F_{ab}\right]}
\ee
where in the above $T_D$ is the D-brane tension, $V$ is the volume of the internal space in which D-branes
may be wrapping. $g_{ab}$ (with $a,b$ the $q+1$ dimensional world-volume indices and the radial
coordinate $r$) is the induced metric and 
$F_{ab}$ is the field-strength of the $U(1)$ gauge field. Since the only non-zero component of the field-strength
is $F_{0r}$, we can use it and perform the integrations over $dt$ and $d^qx$ in \eqref{dbi} to get,
\be\label{dbi1}
S = -{\cal N} \int dr g_{xx}^{q/2}\sqrt{|g_{tt}| g_{rr} - (2\pi\alpha' A_t')^2} 
\ee
where $S$ is the action density $\hat{S}/(\int dt d^qx)$ and ${\cal N} = N_f T_D V$.
Also `prime' denotes the derivative with respect to $r$. Now  
using the metric in \eqref{hvlif} we can solve $A_t'$ from \eqref{dbi1} as,
\be\label{soln}
A_t' = \frac{\hat d}{2\pi\alpha'}\frac{r^{-(z+1-\frac{2\theta}{d})}}{\sqrt{r^{-2q(1-\frac{\theta}{d})}+\hat{d}^2}}
\ee
where $\hat d = (2\pi\alpha'{\cal N})^{-1}\langle J_t\rangle$ is proportional to the charge density. Note
that the background \eqref{hvlif} has been shown in \cite{Dong:2012se} to be stable only when $d-\theta>0$ and we
will assume that in our following calculation. Further it must also satisfy the null energy condition 
\cite{Dong:2012se}
\be\label{NEC}
(d-\theta)(d(z-1)-\theta) \geq 0, \qquad (z-1)(d+z-\theta) \geq 0
\ee
so that it represents a physically sensible gravity dual. As the probe does not backreact to the background
we will use these conditions while obtaining the temperature dependence of specific heat in the next section.

\vspace{.5cm}

\noindent{\it 3. Probe thermodynamics} : In order to understand the thermodynamics we put a black hole
in the background geometry \eqref{hvlif}. We will obtain the low temparture behavior of the specific
heat and this in turn will help us to understand the nature of the matter described by the holographic
model given in the previous section. The background geometry now is given as,
\be\label{hvlifT}
ds^2_{d+2} =  r^{\frac{2\theta}{d}}\left[-f(r)\frac{dt^2}{r^{2z}} + \frac{\sum_{i=1}^d (dx^i)^2}{r^2} + \frac{dr^2}{f(r)r^2}\right]
\ee  
where $f(r) = 1-(r/r_H)^{(d+z-\theta)}$ and $r_H$ denotes the radius of the horizon which is related to the temperature
$T$ of the black hole as $r_H = ((d+z-\theta)/(4\pi T))^{1/z}$. Even with this new geometry the action \eqref{dbi1}
and its solution \eqref{soln} remain the same. However, the integration limit in \eqref{dbi1} will change and it
will start from $r_H$ instead of zero. We evaluate the on-shell action by using the equation of motion \eqref{soln}
in \eqref{dbi1} and this is nothing but the negative Gibbs potential ($\Omega$) in the grand canonical ensemble.
Thus we have
\be\label{potential}
\Omega = -S = {\cal N} \int_{r_H}^{\infty} dr \frac{r^{-(z+1-
\frac{2\theta}{d} +q(1-\frac{\theta}{d}))}}{\sqrt{1 + \hat{d}^2 r^{2q(1-\frac{\theta}{d})}}}
\ee
The Gibbs potential is a function of temperature and the chemical potential, where the chemical potential can be
obtained by integrating $A_t'$ in \eqref{soln}, i.e.,
\be\label{chemical}
\mu = \int_{r_H}^{\infty} dr A_t' = \frac{\hat d}{2\pi\alpha'}\int_{r_H}^{\infty} dr \frac{r^{-(z+1-\frac{2\theta}{d})}}{\sqrt{r^{-2q(1-\frac{\theta}{d})}+\hat{d}^2}}
\ee
Both the integrals in \eqref{potential} and \eqref{chemical} can be performed\footnote{There are some subtleties here. The
integrals actually have UV divergences which are hidden in $\Omega_0$ and $\mu_0$. However, we have given in \eqref{potchem0}
their finite forms assuming that the divergences can be removed by a proper holographic renormalization technique
similar to the ones described for asymptotically AdS space-times in \cite{Bianchi:2001kw, Skenderis:2002wp, Karch:2005ms}.} 
and expressed in terms of hypergeometric functions
as follows,
\bea\label{potchem}
\Omega &=& \Omega_0 - {\cal N} \frac{r_H^{-(2mq + 2m +z -2)}}{\hat{d}(2mq + 2m + z-2)}\, _2F_1\left(\half,\,1+\frac{2m+z-2}{2mq};
\,2 + \frac{2m+z-2}{2mq};\, -\frac{r_H^{-2mq}}{\hat{d}^2}\right)\nn
\mu &=& \mu_0 - \frac{r_H^{-(2m+z-2)}}{2\pi\alpha' (2m+z-2)}\, _2F_1\left(\half,\, \frac{2m+z-2}{2mq};\, 1+\frac{2m+z-2}{2mq};\,
-\frac{r_H^{-2mq}}{\hat{d}^2}\right)
\eea
where we have introduced the parameter $m \equiv 1 - \theta/d$ for brevity. Also, in \eqref{potchem} $\Omega_0$ and $\mu_0$
denote the corresponding quantities at zero temperature and are given as,
\be\label{potchem0}
\mu_0 = \frac{\Gamma\left(\frac{-2m-z+2+mq}{2mq}\right)\Gamma\left(\frac{2m+z-2}{2mq}\right)}{4\pi\alpha' mq\Gamma\left(\half\right)}
\hat{d}^{\frac{2m+z-2}{mq}},\,\,\Omega_0 = -2\pi\alpha'{\cal N}\hat{d}\left(\frac{2m+z-2}{2m+z-2+mq}\right)\mu_0
\ee
Note that since we have $d>\theta$, $m$ must be positive. For $\theta=0$, we have $m=1$ and in that case our results \eqref{potchem},
\eqref{potchem0} match with those given in \cite{HoyosBadajoz:2010kd}. Further for $\theta=0$ and $z=1$, i.e. for the relativistic case our results 
agree with those given in \cite{Karch:2009zz}. Thus we find the Gibbs potential $\Omega$ as a function of temperature $T \sim r_H^{-z}$ and 
chemical potential $\mu$. The Gibbs potential in \eqref{potchem} can be expanded for small temperature ($r_H \to \infty$) and the
result is,
\be\label{smallT}
\Omega = \frac{2\pi\alpha'{\cal N} \hat{d}(2m+z-2)}{2m+z-2+mq}\left[\mu + \frac{(2\pi\alpha')^{-1}}{2m+z-2}
\left(\frac{d+z-\theta}{4\pi}\right)^{\frac{2\theta}{dz}-1} T^{1-\frac{2\theta}{dz}} + O(T^{1-\frac{2\theta}{dz}+\frac{2mq}{z}})\right] 
\ee   
From the first term in \eqref{smallT} we find that varying $\Omega$ with respect to the chemical potential we must get
the charge density and that is proportional to $\hat{d}$ as expected. The second term has a non-trivial temperature dependence
since $2\theta/(dz)$ can not equal 1 as this will violate the first null energy condition given in \eqref{NEC}. 
Varying the second term with respect to $T$, we get an expression of entropy density and it has the form,
\be\label{entropy}
s = -\left(\frac{\partial \Omega}{\partial T}\right)_{\mu} = \frac{{\cal N}\hat{d}}{2m+z-2+mq}\left(\frac{d+z-\theta}{4\pi}\right)^{\frac{2\theta}{dz}-1}
\left(1-\frac{2\theta}{dz}\right)T^{-\frac{2\theta}{dz}}+O(T^{-\frac{2\theta}{dz}+\frac{2mq}{z}})
\ee  
We thus find from \eqref{entropy} that when there is no hyperscaling violation (for $\theta=0$), the entropy density is constant
at zero temperature as noted in \cite{Karch:2009zz, HoyosBadajoz:2010kd}. However, for non-zero hyperscaling it can vanish if $\theta/z$ 
is negative (or it
can blow up if $\theta/z$ is positive, which indicates instability and we exclude this case). The specific heat at low temperature 
can also be calculated from \eqref{entropy} by varying it with respect to $T$ and has the form $c_V = T(\partial s/\partial T) 
\sim -2\theta/(dz) T^{-2\theta/(dz)}$. Again we find that the specific heat is positive, i.e., the system is stable as long as
$\theta/z < 0$. This implies that for the stability of the system we must have either (i) $\theta >0$, $z<0$ or (ii) $\theta <0$,
$z>0$. It can be easily checked that for case (i), the first of the null energy conditions \eqref{NEC} is violated and therefore
this case does not give sensible gravity dual. For case (ii) both the null energy conditions can be satisfied as long as $z \geq 1$.
We thus conclude that our gravity dual is sensible and stable only when $\theta < 0$ and $z\geq 1$. As $\theta<0$, we can write $c_V 
\sim 2|\theta|/(dz) T^{2|\theta|/(dz)}$. Thus $c_V$ scales linearly with temperature or the system behaves like Fermi liquid if
$2|\theta|=dz$ and $c_V$ scales as $T^q$ or the system behaves like Bose liquid if $2|\theta| = dzq$. When the parameters
do not satisfy these two conditions, i.e., for the generic values of the parameters the system behaves like neither a Fermi liquid
nor a Bose liquid but a new kind of quantum liquid.

\vspace{.5cm}

\noindent{\it 4. Zero sound} : The zero sound mode appears at zero temperature as a pole in the Fourier transformed retarded
two point
function of the density operator \cite{Lifshitz, Abrikosov, Pines, Negele}. On the gravity side the pole arises as the 
quasinormal frequency of the background \cite{Son:2002sd, Kovtun:2005ev}.
As in this case we are interested in the density or current operator, it is sufficient to consider fluctuations of the $U(1)$
field of the background only with non-trivial background component $A_t$. As the field theory is isotropic the fluctuations
can be chosen to depend on $r$, $t$ and one of $x^i$'s which we call $x$ and so\footnote{Fluctuations in holographic theories with
hyperscaling violation has been studied in \cite{Edalati:2012tc}.},
\be\label{fluc}
A_\mu(r) \to A_\mu (r) + a_\mu(r,t,x)
\ee
Using this into the DBI action \eqref{dbi1} and expanding it upto quadratic order in fluctuations we obtain (in the $a_r=0$ gauge)
[\cite{HoyosBadajoz:2010kd}, Here and in some of the following equations we use their notation and general formalism],
\be\label{action2}
S^{(2)} = \frac{{\cal N}}{2}\int dr dt dx g_{xx}^{q/2}\left[\frac{g_{rr}f_{tx}^2 - |g_{tt}|(2\pi\alpha'a_x')^2}{g_{xx}\sqrt{|g_{tt}|g_{rr} -
(2\pi\alpha' A_t')^2}} + \frac{|g_{tt}|g_{rr}(2\pi\alpha'a_t')^2}{\left(|g_{tt}|g_{rr} - (2\pi\alpha' A_t')^2\right)^{3/2}}\right]
\ee      
where $f_{tx} = 2\pi\alpha'(\partial_t a_x-\partial_x a_t)$. Introducing Fourier components 
\be\label{fourier}
a_\mu (r,t,x)=\frac{1}{(2\pi)^2}\int d\omega dk e^{-i\omega t +i k x} a_\mu (r,\omega, k)
\ee 
the equations of motion for the fluctuations take the form,
\bea\label{eom}
& & \frac{d}{dr}\left[\frac{g_{xx}^{q/2}|g_{tt}|g_{rr}a_t'}{\left(|g_{tt}|g_{rr}-(2\pi\alpha'A_t')^2\right)^{3/2}}\right]
-\frac{g_{xx}^{q/2-1}g_{rr}}{\sqrt{|g_{tt}|g_{rr}-(2\pi\alpha'A_t')^2}}\left(k^2a_t + \omega k a_x\right)\,\, =\,\, 0\nn
& & \frac{d}{dr}\left[\frac{g_{xx}^{q/2-1}|g_{tt}|a_x'}{\sqrt{|g_{tt}|g_{rr}-(2\pi\alpha'A_t')^2}}\right]
+\frac{g_{xx}^{q/2-1}g_{rr}}{\sqrt{|g_{tt}|g_{rr}-(2\pi\alpha'A_t')^2}}\left(\omega^2 a_x + \omega k a_t\right)\,\, =\,\, 0
\eea
There is a constraint arising from the equation of motion of $a_r$ in the $a_r=0$ gauge and is given as,
\be\label{constraint}
g_{rr}g_{xx}\omega a_t' + \left(|g_{tt}|g_{rr}-(2\pi\alpha'A_t')^2\right)ka_x' = 0
\ee
Using the constraint equation and the first equation in \eqref{eom} (as the second equation in \eqref{eom}
can be obtained from the constraint and the first equation), we can write an equation for the gauge invariant
field $E(r,\omega,k) = \omega a_x + k a_t$ as,
\be\label{eomE}
E'' + \left[\frac{d}{dr}\ln\left(\frac{g_{xx}^{(q-3)/2}g_{rr}^{-1/2}|g_{tt}|}{u(k^2u^2-\omega^2)}\right)\right]E'
-\frac{g_{rr}}{|g_{tt}|}(k^2u^2-\omega^2)E \,\,=\,\,0
\ee
where 
\be\label{udefn}
u(r)\,\,=\,\, \sqrt{\frac{|g_{tt}|g_{rr} - (2\pi\alpha'A_t')^2}{g_{rr}g_{xx}}} \,\,=\,\, \sqrt{\frac{|g_{tt}|}{g_{xx}
\left(1+\hat{d}^2 g_{xx}^{-q}\right)}}\,\,=\,\, \frac{r^{-(z-1)}}{\sqrt{1+\hat{d}^2 r^{2mq}}}
\ee
The quadratic action for the fluctuations $S^{(2)}$ (in \eqref{action2}) can be written in terms of $E$ as,
\be\label{action3}
S^{(2)}\, \,=\,\, (2\pi\alpha')^2\frac{{\cal N}}{2}\int dr d\omega dk \frac{g_{xx}^{(q-3)/2}g_{rr}^{1/2}}{u}
\left[E^2 + \frac{|g_{tt}|}{g_{rr}(u^2k^2-\omega^2)}E'^2\right]
\ee
Introducing a cut-off near the boundary at $r=\epsilon$, using equation of motion \eqref{eomE}, and integrating
by parts we obtain the action in the limit $\epsilon \to 0$,
\be\label{action4}
S^{(2)}\,\,=\,\, -(2\pi\alpha')^2 \frac{{\cal N}}{2}\int d\omega dk \frac{\epsilon^{2m+z-1-mq}}{k^2} E(\epsilon)E'(\epsilon) 
\ee
Now the strategy to obtain the retarded two point function as explained in \cite{Karch:2009zz}, is to first solve 
the equation of motion
\eqref{eomE} for $E$ using the incoming boundary condition at the horizon $r \to \infty$ and then substitute it
into \eqref{action4} and finally take functional derivative of the action with respect to $a_t$. Thus we have,
\be\label{funcder}
G_R^{tt}(\omega, k)\,\,=\,\, \frac{\delta^2}{\delta a_t(\epsilon)^2}S^{(2)}\,\,=\,\, \left(\frac{\delta E(\epsilon)}
{\delta a_t(\epsilon)}\right)^2 \frac{\delta^2}{\delta E(\epsilon)^2}S^{(2)}
\ee
Since $S^{(2)}$ is a function of $E$, we will obtain the low frequency and low momentum form of 
$\Pi(\omega, k) \equiv (\delta^2/\delta E^2)S^{(2)}$ and look at its 
pole structure. Now in general it is difficult to solve $E$ from \eqref{eomE} and so, to obtain $\Pi(\omega, k)$ we will
solve the equation \eqref{eomE} in two different limits and then match the two solutions in the overlapping region 
\cite{Karch:2009zz, HoyosBadajoz:2010kd}.
To be precise, we first obtain a solution when $r$ is very large and then take small frequency and small momentum
limit, which means $\omega r^z \ll 1$ and $kr\ll 1$, with $\omega k^{-z}$ fixed. Then we reverse the process, i.e., 
take the small frequency, small momentum limit first, solve the equation and then take $r$ very large.              

At large $r$, the equation for $E$, \eqref{eomE}, reduces to,
\be\label{eomE1}
E'' + \frac{2m-(z-1)}{r} E' + \omega^2 r^{2(z-1)}E\,\,=\,\, 0
\ee
where we have used the background metric \eqref{hvlif}. This equation has a solution with incoming wave boundary 
condition in terms of a Hankel fucntion 
$E(r) = C (\omega r^z/(2z))^{1/2 - m/z} H^{(1)}_{1/2-m/z}(\omega r^z/z)$, where $C$ is a constant. Now we take small frequency,
small momentum limit, $\omega r^z \ll 1$. In this limit $E(r)$ takes two different forms depending on the value of $z$.
Thus, 
\bea\label{solution1a}
E &\approx & C + C \frac{2i}{\pi}\left[\log\left(\omega r^{2m}\right) - \log(4m) + \gamma\right], \qquad\qquad {\rm for}\,\, z=2m\\
\label{solution1b}
E &\approx & \tilde{C} \Gamma\left(\frac{m}{z} + \half\right)^{-1}\left[1 - \tan \frac{m\pi}{z}\right] - i \frac{\tilde{C}}{\pi}
\Gamma\left(\frac{m}{z} - \half\right)\left(\frac{\omega r^z}{2z}\right)^{1-\frac{2m}{z}}, \quad {\rm for}\,\, z \neq 2m
\eea 
Note in the above that $m = 1 + |\theta|/d$ as defined before. Also, $\tilde{C}$ is another constant and is related to $C$ by
$\tilde{C} = -iCe^{-im\pi/z}\cos(m\pi/z)$. $\gamma$ is the Euler-Mascheroni number. 

On the other hand, in the small frequency, small momentum limit ($\omega r^z \ll 1$, $kr\ll 1$, with $\omega k^{-z}$ fixed), the
equation \eqref{eomE} takes the form,
\bea\label{eomE2}
& & E'' + \frac{1}{r}\Big[(2-q)m - 2(z-1)\Big.\nn
& & \qquad\qquad\qquad \Big. + \left((z-1) + \frac{qm \hat{d}^2 r^{2qm}}{1 + \hat{d}^2 r^{2qm}}\right)\left(\frac{3k^2r^2
- \omega^2 r^{2z}(1+\hat{d}^2r^{2qm})}{k^2r^2 -  \omega^2 r^{2z}(1+\hat{d}^2r^{2qm})}\right) \Big]E' = 0
\eea
where again we have used the background metric \eqref{hvlif}. This equation \eqref{eomE2} has solution in terms of
hypergeometric functions and is given as,
\bea\label{solutioneom2}
& & E(r) = \nn
& & C_1 + C_2\left[\frac{k^2 r^{2+m(q-2)-z}}{2+m(q-2)-z}\, _2F_1\left(\frac{3}{2}, \frac{2+m(q-2)-z}{2mq}; 1+ \frac{2+m(q-2)-z}{2mq};
-\hat{d}^2 r^{2mq}\right)\right.\nn
& & \qquad\qquad \left. -\frac{\omega^2 r^{m(q-2)+z}}{m(q-2)+z}\, _2F_1\left(\frac{1}{2}, \frac{m(q-2)+z}{2mq}; 1+ \frac{m(q-2)+z}{2mq};
-\hat{d}^2 r^{2mq}\right)\right]
\eea 
where $C_1$ and $C_2$ are two constants. Now taking $r \to \infty$, $E(r)$ takes two different forms depending on the value of $z$.
They are given as,   
\bea\label{solution2a}
E &\approx& C_1 + \frac{C_2}{\hat{d}}\Big[\frac{k^2(2m-1)}{(mq)^2} B\left(\frac{-2+4m}{2mq},\frac{2-4m}{2mq}+\half\right)
\hat{d}^{\frac{-2+4m}{mq}}\Big.\nn
& & \Big.- \frac{\omega^2}{mq} \log\left(\frac{2\hat{d}}{\omega^{q/2}}\right)- \frac{\omega^2}{2m}\log\left(\omega r^{2m}\right)\Big], 
\qquad\qquad {\rm for}\,\, z=2m\\
\label{solution2b}
E &\approx& C_1 + \frac{C_2}{\hat{d}}\Big[\frac{k^2(2m+z-2)}{2(mq)^2} B\left(\frac{2m+z-2}{2mq},\frac{-2m-z+2}{2mq}+\half\right)
\hat{d}^{\frac{2m+z-2}{mq}}\Big.\nn
& & \Big.- \frac{\omega^2}{2mq} B\left(\frac{2m-z}{2mq},\frac{z-2m}{2mq}+\half\right)\hat{d}^{\frac{2m-z}{mq}}  - 
\frac{\omega^2}{z-2m}r^{z-2m}\Big], 
\quad {\rm for}\,\, z\neq 2m 
\eea
where $B(x,y)$ is the Beta function. Comparing the solutions \eqref{solution1a} with \eqref{solution2a} and also comparing the 
solutions \eqref{solution1b} with \eqref{solution2b} we find that they have the same structure. When $z=2m$, both the solutions
have a log term and a constant term and when $z\neq 2m$, they have a $r^{z-2m}$ term and a constant term. This actually 
enables us to eliminate one of the constants $C$ in \eqref{solution1a} (or $\tilde{C}$ in \eqref{solution1b}) and relate constants
$C_1$ in terms of $C_2$ (both in \eqref{solution2a} and \eqref{solution2b}). This will be needed, as we will see, to evaluate the 
action \eqref{action4} as well as the two point function \eqref{funcder} or equivalently $\Pi(\omega, k)$. To evaluate the action
\eqref{action4} we need the boundary behavior of $E(r)$ and this can be obtained from the general small frequency, small momentum 
solution \eqref{solutioneom2} as $r\to 0$. In fact one can check from \eqref{solutioneom2} that in the small $r$ limit $E(r)$ and
$E'(r)$ behave as (for $z>1$),
\bea\label{eeprime}
E &\approx& C_1 + C_2 \frac{k^2 r^{2+m(q-2)-z}}{2 + m(q-2) - z}\nn
E' &\approx& C_2 k^2 r^{2+m(q-2)-z-1}
\eea
For $z+2m>2+mq$, the second term will go to zero for $r \to 0$ and so $E \approx C_1$. However, as it has been argued in 
\cite{HoyosBadajoz:2010kd},
even for $z+2m\leq 2+mq$, $E$ can be approximated as $C_1$. The reason is that in the latter case one can add a suitable boundary
term in the action which does not affect the equation of motion but cancels the divergence as we take $r\to 0$. The action $S^{(2)}$
in that case changes a sign which is not important for the pole structure. So, using $E \approx C_1$ and $E'$ from above the action
\eqref{action4} takes the form,
\be\label{action5}
S^{(2)}\,\,=\,\, -(2\pi\alpha')^2 \frac{{\cal N}}{2}\int d\omega dk \frac{\epsilon^{2m+z-1-mq}}{k^2} E(\epsilon)E'(\epsilon)\,\,=\,\,
 -(2\pi\alpha')^2 \frac{{\cal N}}{2}\int d\omega dk C_1 C_2 
\ee   
Now $\Pi(\omega, k)$ can be calculated as follows,
\be\label{picalc}
\Pi(\omega, k) = \left.\frac{\delta^2}{\delta E(\epsilon)^2}S^{(2)}\right|_{\epsilon \to 0} =  
\left.\frac{\delta^2}{\delta C_1^2}S^{(2)}\right|_{\epsilon \to 0} =
 \left. \frac{dC_2}{dC_1}\frac{\delta^2}{\delta C_1 \delta C_2}S^{(2)}\right|_{\epsilon \to 0} =
-\frac{{\cal N}}{2}(2\pi\alpha')^2 \frac{dC_2}{dC_1} 
\ee
Using \eqref{picalc} and the obtained relations between $C_1$ and $C_2$ as indicated before, we get the low
frequency, low momentum form of $\Pi(\omega, k)$ as,
\be\label{pi}
\Pi(\omega, k) = \frac{P {\cal N}}{\alpha_1 k^2 - \alpha_2 \omega^2 - \alpha_3 G_0(\omega)}
\ee
where $P$, $\alpha_1$, $\alpha_2$, $\alpha_3$ are some constants and $G_0$ is a function of $\omega$. For
$z=2m=2(1+|\theta|/d)$, they have the forms,
\bea\label{zeq2mvalues}
P &=& 8 \pi m \hat{d} \alpha'^2\nn
\alpha_1 &=& \frac{4m(2m-1)}{\pi (mq)^2}\hat{d}^{\frac{4m-2}{mq}} B\left(\frac{4m-2}{2mq},\frac{2-4m}{2mq}+\half\right)\nn
\alpha_2 &=& i\nn
\alpha_3 &=& -\frac{1}{\pi}\nn
G_0(\omega) &=& \omega^2 \log(\alpha \omega^2)
\eea
where $\alpha = e^{2\gamma} (2\hat{d})^{-4/q}/16$ is a constant, and for $z \neq 2m = 2(1+|\theta|/d)$, they have
the forms,
\bea\label{zneq2mvalues}
P &=& 2\pi\alpha'^2 \hat{d} (2m-z)(2z)^{\frac{2m}{z}-1}\Gamma\left(\frac{m}{z}+\half \right)\Gamma\left(\frac{m}{z}-\half \right)\nn
\alpha_1 &=& \frac{(2m-z)(2m+z-2)(2z)^{\frac{2m}{z}-1}}{2(mq)^2\pi\hat{d}^{\frac{2-2m-z}{mq}}} 
\Gamma\left(\frac{m}{z}+\half \right)\Gamma\left(\frac{m}{z}-\half \right)\nn
& &\qquad\qquad\qquad \times B\left(\frac{2m+z-2}{2mq}, \frac{2-2m-z}{2mq}+\half\right)\nn
\alpha_2 &=&  \frac{(2m-z)(2z)^{\frac{2m}{z}-1}}{2mq\pi\hat{d}^{\frac{z-2m}{mq}}} 
\Gamma\left(\frac{m}{z}+\half \right)\Gamma\left(\frac{m}{z}-\half \right) B\left(\frac{2m-z}{2mq}, \frac{z-2m}{2mq}+\half\right)\nn  
\alpha_3 &=& i + \tan{\frac{\pi m}{z}}\nn
G_0(\omega) &=& \omega^{1+\frac{2m}{z}}
\eea 
It is clear from \eqref{pi} that $\Pi(\omega, k)$ and so, the two point retarded correlation function will have a pole when
the denominator vanishes, i.e., when
\be\label{komega}
k(\omega) = \pm \frac{1}{\sqrt{\alpha_1}}\sqrt{\alpha_2 \omega^2 + \alpha_3 G_0(\omega)}
\ee
From the expression of $G_0(\omega)$ in \eqref{zneq2mvalues}, we find that when $1 \leq z < 2m$, $\omega^2$ term dominates
$G_0(\omega)$ term and when $z > 2m$, $G_0(\omega)$ term will dominate $\omega^2$ term. So, the pole structure will be different
for these two cases. Let us first consider $1 \leq z < 2m$. In this case we can write \eqref{komega} in the small frequency as,
\be\label{komega1}
k(\omega) = \pm \omega \sqrt{\frac{\alpha_2}{\alpha_1}} \left[1+\frac{\alpha_3}{2\alpha_2}\omega^{\frac{2m}{z}-1} + O(\omega^{\frac{4m}{z}-2})\right]
\ee
which can then be inverted to give,
\be\label{omegak1}
\omega(k) = \pm\sqrt{\frac{\alpha_1}{\alpha_2}} k - \frac{\alpha_3}{2\alpha_2}\left(\frac{\alpha_1}{\alpha_2}\right)^{\frac{m}{z}} k^{\frac{2m}{z}} + 
O(k^{-1+\frac{4m}{z}})
\ee
We note from \eqref{zneq2mvalues} that when $z \neq 2m$, $\alpha_1$ and $\alpha_2$ are real whereas $\alpha_3$ is complex. Therefore, the first
term in \eqref{omegak1} is real and the second term gives an imaginary contribution. Now since the imaginary part goes as $k^{2m/z}$ which is smaller
than the real part (which goes as $k$) for small $k$ and $1\leq z <2m$, this mode behaves like a quasiparticle. Also, when $z = m = 1 +|\theta|/d$, 
the imaginary part goes as $k^2$, exactly like Fermi liquid. We found in section 3, from the behavior of specific heat with temperature that when
$2|\theta| = dz$, it has a linear or Fermi liquid like behavior. So, combining these two we find that for $z=2$, both the specific heat and the
zero sound behave like Fermi liquid. When $2|\theta| = dz$ (or $2|\theta| = dqz$) and $z \neq m = 1+|\theta|/d$, the specific heat behaves like
Fermi (or Bose) liquid, but the zero sound does not behave like Fermi liquid and when $2|\theta| \neq dz$ but $z=m=1+|\theta|/d$, the specific heat
does not behave like Fermi liquid, but the zero sound does. However, in this range of $z$ not satisfying either of the conditions, there exits a zero
sound although the system behaves like a new kind of quantum liquid.     

The speed of the zero sound can be found from the real part of the dispersion relation \eqref{omegak1} and is given as,
\be\label{speed}
v_0^2 = \frac{\alpha_1}{\alpha_2} = \frac{2m+z-2}{mq}\hat{d}^{\frac{2(z-1)}{mq}}\frac{\Gamma\left(\frac{2m+z-2}{2mq}\right)
\Gamma\left(\half - \frac{2m+z-2}{2mq}\right)}
{\Gamma\left(\frac{2m-z}{2mq}\right)\Gamma\left(\half - \frac{2m-z}{2mq}\right)}
\ee
When $1 < z < 2m$, the finiteness or the vanishing of the velocity is determined by the poles of the gamma functions $\Gamma(1/2 - (2m+z-2)/(2mq))$
and $\Gamma((2m-z)/(2mq))$ as $z \to 2m$, the details of which will depend on the values of $m$ and $q$. In particular, when $q=4-2/m$ (note that since
$q$ is an integer $m$ can not take any value), all the $\Gamma$ functions with their poles cancel and we get $v_0^2 = (2m+z-2)/(4m-2) \hat{d}^{(z-1)/(2m-1)}$
which takes a finite value $\hat{d}$ as $z \to 2m$. Note that when $\theta =0$, i.e., $m=1$ the above expression for zero sound speed
\eqref{speed} matches with that given in \cite{HoyosBadajoz:2010kd} and in addition when $z=1$, it matches with that given in 
\cite{Karch:2009zz}.

When $z>2m$, $G_0(\omega)$ will dominate $\omega^2$ term for small $\omega$ in \eqref{komega}. In that case $k(\omega)$ can be written as, 
\be\label{komega2}
k(\omega) = \pm \sqrt{\frac{\alpha_3}{\alpha_1}}\omega^{\frac{z+2m}{2z}}\left[1+\frac{\alpha_2}{2\alpha_3}\omega^{1-\frac{2m}{z}} + O(\omega^{2-\frac{4m}{z}})\right]
\ee
which can be inverted to give the following dispersion relation,
\be\label{omegak2}
\omega(k) = \left(\frac{\alpha_1}{\alpha_3}\right)^{\frac{z}{z+2m}} k^{\frac{2z}{z+2m}} - \frac{z}{z+2m} \left(\frac{\alpha_2}{\alpha_3}\right)
\left(\frac{\alpha_1}{\alpha_3}\right)^{\frac{2(z-m)}{z+2m}} k^{\frac{4(z-m)}{z+2m}} + O(k^{\frac{2(3z-4m)}{z+2m}})
\ee
As we have seen in \eqref{zneq2mvalues}, in this case, $\alpha_1$, $\alpha_2$ are real and $\alpha_3$ is complex. So, both the real and
imaginary parts come from the leading term, i.e., they have the same order and therefore this is not a quasiparticle. There is no zero
mode for $z>2m=2(1+|\theta|/d)$.

When $z=2m$, using the form of $G_0(\omega)$ from \eqref{zeq2mvalues} into \eqref{komega}, the small frequency expansion can be
written as,
\be\label{komega3}
k(\omega) = \pm \frac{\omega}{\sqrt{\alpha_1}}\left[\sqrt{\alpha_3} [\log(\alpha\omega^2)]^{\half} + \frac{\alpha_2}{2\sqrt{\alpha_3}}
[\log(\alpha\omega^2)]^{-\half} + O([\log(\alpha\omega^2)]^{-\frac{3}{2}})\right].
\ee 
We note that because of the logarithmic terms it is not possible to invert the relation \eqref{komega3} to obtain $\omega(k)$ completely.
At best we can express $\omega [\log (\alpha \omega^2)]^{1/2}$ as
\be\label{omegak3}
\omega(k) [\log (\alpha \omega^2(k))]^{\half} = \pm \sqrt{\frac{\alpha_1}{\alpha_3}}k - \frac{\alpha_2 \sqrt{\alpha_1}}{2 \alpha_3^{\frac{3}{2}}}
\frac{k}{[\log(\alpha \omega^2)]} + O\left(\frac{k}{\alpha_3^{\frac{5}{2}}[\log(\alpha\omega^2)]^{2}}\right)
\ee
where the right hand side also involves $\omega$ in the form of $\log(\alpha \omega^2)$.
Now as $\omega \to 0$, $\log(\alpha \omega^2) \to -\infty$ and therefore the leading contribution to $\omega \sqrt{\log(\alpha \omega^2)}$,
as can be seen from \eqref{omegak3}, is linear in $k$. However, since this dispersion relation differs from the usual zero sound mode dispersion  
relation by logarithmic terms we conclude that it does not represent a quasiparticle and there is no zero sound mode at $z=2m=2(1+|\theta|/d)$.

\vspace{.5cm}

\noindent{\it 5. AC conductivity} : AC conductivity of the system can be found from the two-point retarded current correlator by the Kubo's
formula as,
\be\label{sigma}
\sigma(\omega) = -\frac{i}{\omega} G_R^{xx}(\omega, k=0)
\ee
The current correlator can be obtained from the quadratic action $S^{(2)}$ as,
\be\label{correlator}
G_R^{xx}(\omega, k) = \frac{\delta^2}{\delta a_x(\epsilon)^2}S^{(2)} = \left(\frac{\delta E(\epsilon)}{\delta a_x}\right)^2 
\frac{\delta^2}{\delta E(\epsilon)^2} S^{(2)} = \omega^2 \Pi(\omega, k)
\ee
So, once we have $\Pi(\omega, k)$ it is trivial to write $G_R^{xx} (\omega, k)$ and from there we can obtain $\sigma(\omega)$ 
at small frequency as,
\be\label{sigma0} 
\sigma(\omega) = -\frac{i}{\omega} G_R^{xx}(\omega, k=0) = -i\omega \Pi(\omega, k=0) \xrightarrow{\omega \rightarrow 0}
i P {\cal N} 
\begin{cases}
\alpha_2^{-1} \omega^{-1}, & z<2m\\
\alpha_3^{-1} \omega^{-1} [\log(\alpha \omega^2)]^{-1}, &  z=2m\\
\alpha_3^{-1} \omega^{-\frac{2m}{z}}, &  z>2m
\end{cases}
\ee
where we have used the form of $\Pi(\omega,k)$ given in \eqref{pi}. The constants are given in \eqref{zeq2mvalues} for $z=2m$
and in \eqref{zneq2mvalues} for $z \neq 2m$. We thus find that for $z<2m=2(1+|\theta|/d)$, the AC conductivity behaves as $\sim
i/\omega$. This is the standard Drude conductivity, i.e., the high-frequency limit of Drude model. This result is quite universal
as it was found true even without hyperscaling violation ($\theta=0$) \cite{HoyosBadajoz:2010kd} and also in 2+1 dimensional 
system \cite{Hartnoll:2009ns}, only the range of $z$
in which this behavior occurs depends on the hyperscaling violation exponent and the dimension of the system. On the other hand,
when $z>2m$, the AC conductivity has an anomalous power-law tails as also found in \cite{Hartnoll:2009ns} and 
\cite{HoyosBadajoz:2010kd}. In some 2+1 dimensional
system the exponent has been found to be 0.65, which can occur in this case if $z=3m=3(1+|\theta|/d)$.  

\vspace{.5cm}

\noindent{\it 6. Conclusion} : In this paper we have modelled strange metals with hyperscaling violation by introducing
probe D-branes in the background of Lifshitz space-time with hyperscaling violation. The behavior of the specific heat
and the null energy condition dictate that the system must have $\theta<0$ and $z\geq 1$ in order to have a sensible and 
stable gravity dual. The specific heat of this system shows 
in general a new kind of quantum liquid behavior, $c_V \sim T^{2|\theta|/(dz)}$. In particular,
when $2|\theta|=dz$, $c_V \sim T$, the system behaves like Fermi liquid and when $2|\theta|=dqz$, where $q$ is the spatial dimension
of the boundary theory, $c_V \sim T^q$, the system behaves like Bose liquid. We have then studied the fate of the zero sound
in this system which were known to exist for strange metals without hyperscaling violation. We found that even though the system
in general behaves like a non-Fermi liquid, there always exists a quasiparticle for the zero sound mode as long as $z$ satisfies
$1 \leq z<2m=2(1+\theta|/d)$. When $\theta$ vanishes, we recover the results of ref.\cite{HoyosBadajoz:2010kd}. However, the zero mode 
does not survive
when $z$ goes outside this range, i.e., for $z \geq 2m = 2(1+|\theta|/d)$. In the specified range of $z$, when zero sound mode exists
we found that the real part is linear in $k$ and from there we found the speed of the zero sound. The imaginary part of the dispersion
curve in general goes as $k^{2m/z}$ and so it behaves like a Fermi liquid when $z=m=1+|\theta|/d$. Actually we found that both the
specific heat and the zero sound behave like Fermi liquid for $z=2$. We have also discussed various cases when specific heat behaves like
Fermi or Bose liquid and zero sound behaves like Fermi liquid or non-Fermi liquid. We have also obtained the form of AC conductivity
in this system. Here we found that for $z<2m$, the AC conductivity goes as $i/\omega$ which is nothing but the high frequency limit
of the Drude model also known as Drude conductivity. On the other hand, for $z>2m$, AC conductivity shows an anomalous power law tail
and for $z=3m$ the power becomes 0.65 as has been found for some 2+1 dimensional system.

In our model we have assumed that the background Lifshitz space-time with hyperscaling violation is obtained from some brane configuration
at large $N$. Such brane configurations have been obtained from string theory in \cite{Dey:2012tg}. However, in all these cases the hyperscaling
violation exponents were found to be positive, whereas, for the background we consider here hyperscaling violation exponents are negative. 
How to obtain such backgrounds from a fundamental theory like string/M theory remains a challenge.

\vspace{.5cm}

\noindent{\it Note} : While preparing this manuscript a paper \cite{Pang:2013ypa} appeared in the arXiv which has 
substantial overlap with
this work. However, their background is a special case of ours, namely, they have $\theta \to -\infty$, $z \to \infty$,
with $\theta/z$ = fixed (negative value). In our case we have $\theta < 0$ and $z \geq 1$.
       
\vspace{.5cm}

\noindent{\it Acknowledgements} :
One of the authors (PD) would like to acknowledge thankfully the financial
support of the Council of Scientific and Industrial Research, India
(SPM-07/489 (0089)/2010-EMR-I). We would like to thank the anonymous referee
for pointing out an error in an earlier version of this paper.

\end{document}